\documentclass[aps,pre,twocolumn]{revtex4-2}
\usepackage{amssymb}
\usepackage{amsmath,bm}
\usepackage{graphicx,color}
\usepackage{bbm}
\usepackage[bottom]{footmisc}
\usepackage{multirow}
\usepackage{xcolor}
\usepackage{xpatch} 
\makeatletter   
\usepackage{xpatch} 

\newcommand{\B}[1]{{\bm{#1}}}

\newcommand{\C}[1]{{\mathcal{#1}}}

\begin{document}
\title{Selection Principle for the Screening Parameters in the Mechanical Response of Amorphous Solids}

\author{Pawandeep Kaur, Itamar Procaccia$^{1}$ and Tuhin Samanta}
\affiliation {Dept. of Chemical Physics, The Weizmann Institute of Science, Rehovot 76100, Israel, $^1$Sino-Europe Complexity Science Center, School of Mathematics, North University of China, Shanxi, Taiyuan 030051, China.}

\begin{abstract}
The mechanical response of amorphous solids to external strains is riddled with plastic events that create topological charges in the resulting displacement field. It was recently shown that the latter leads to screening phenomena that are accompanied by the breaking of both translational and Chiral symmetries. The screening effects are quantified by two screening parameters $\kappa_e$ and $\kappa_o$, which are inverse characteristic lengths that do not exist in classical elasticity. The screening parameters (and the associated lengths) are emergent, and it is important to understand how they are selected. This Letter explores the mechanism of selection of these characteristic lengths in two examples of strain protocols that allow analytic scrutiny. 
\end{abstract}
\maketitle

{\bf Introduction:} The term ``Amorphous solids" spans a wide variety of materials, from silicate and metallic glasses, granular matter like sand or powders, polymeric glasses, gels, and foams, up to biological tissues \cite{08Zal,98Ale,jaeger1996granular,10.1093/pnasnexus/pgae315, BlumPRL,NagelPRB}. Being solid, one might expect that the well-developed classical elasticity theory \cite{Landau} might suffice to describe the responses of such materials to external loads. It turns out however that amorphous solids tend to exhibit plastic responses, and these are prevalent (in the thermodynamic limit) for any amount of external stress \cite{10KLP,11HKLP}. It is well known by now that plastic responses tend to appear with quadrupolar symmetry in the displacement field, and these are known in the jargon as ``Eshelby events" in light of Eshelby's pioneering study of the response to a quadrupolar strain in an elastic material \cite{54Esh}. It is less well known that these plastic responses can lead to screening of the elastic fields.  This
finding is based on recent research, in which it was discovered that the prevalence of plastic events in amorphous solids results in screening phenomena that are akin, but richer and different, to screening effects in electrostatics \cite{21LMMPRS,22MMPRSZ,22BMP,22KMPS,23CMP,24JPS}. Plastic events, which are typically quadrupoles in the displacement field, can act as screening charges. It was shown that when the field of plastic quadrupoles, $\B Q$, is uniform (small gradients), their effect is limited to renormalizing the elastic moduli, but the structure of (linear) elasticity theory remains intact. This is analogous to dipole screening in dielectrics, and we refer to this situation as ``quasi-elastic". On the other hand, when the quadrupole density
becomes high and non-uniform, the presence of effective dipoles (defined by the gradients of the quadrupolar density, ${\C P}^\alpha\equiv \partial _\beta Q^{\alpha\beta}$) cannot be neglected. As in many other areas of statistical physics (e.g. Hexatic \cite{79NH,80ZHN} and Kosterlitz-Thouless transitions \cite{16Kos}, etc.), the presence of dipoles changes the analytic form of the response
to strains, in ways that are in fundamental clash with standard elasticity theory.
This is analogous to Debye screening due to monopoles (charges)  in electrostatics.  We refer to such responses as ``anomalous". It was concluded that one needs to consider a new theory, and this
emergent theory was tested by comparing its predictions to results of extensive experiments and simulations \cite{21LMMPRS,22MMPRSZ,22BMP,22KMPS,23CMP}. 

While dipole screening was observed in both two and three dimensions, in this Letter, we focus on two-dimensional systems in which one can demonstrate a clear transition between quasi-elastic and anomalous response by observing the displacement field that results from a chosen strain protocol. In a purely elastic sheet, the displacement field that arises in a response to a controlled strain satisfies the equation
\begin{equation}
	 \Delta \mathbf{d} + \left (1+ \tilde\lambda \right) \nabla \left(\nabla\cdot \mathbf{d}\right) = 0 \ , \quad\text{purely elastic.}
	\label{d1}
\end{equation}
with the appropriate boundary conditions. Here, 
$\tilde{\lambda} \equiv \frac{\lambda}{\mu}$, where \(\lambda\) and \(\mu\) are the classical Lame' coefficients.
It was shown in a recent series of papers \cite{21LMMPRS,22MMPRSZ,22BMP,22KMPS,23CMP} that in the presence of plastic events that are typical to the response of amorphous solids, the equation for the displacement field changes, to take into account the screening effects that result from plasticity. The equation reads
\begin{equation}
	  \Delta \mathbf{d} + \left (1+\tilde \lambda \right) \nabla \left(\nabla\cdot \mathbf{d}\right) =  -\B \Gamma \mathbf{d}
	\, \quad\text{with screening.}
	\label{d2}
\end{equation}
Here, $\B \Gamma$ is a tensor containing screening parameters that must be specified.
In previous work, it was assumed that in isotropic and homogeneous amorphous solids one can assume that 
$\Gamma$ is diagonal, $ \B \Gamma = \begin{pmatrix}
	\kappa^2 & 0 \\
	0 & \kappa^2
\end{pmatrix}$, leading to an equation to be solved of the form  
\begin{equation}\label{L}
	\Delta \B {d} + (1+\tilde\lambda )\B \nabla (\B \nabla \cdot \B {d}) +k^2\B {d} =0.
\end{equation}
The term $k^2\B {d}$ is responsible for translational symmetry breaking, the introduction of a typical length scale $\ell$, 
$\ell \sim k^{-1}$, and to screening phenomena that change dramatically the expected displacement field $\B d$ from the predictions of Eq.~(\ref{d1}). While experiments and simulations provided support for the predictions of Eq.~(\ref{L}), there was one consequence that was at odds with the data. This is a resulting constitutive equation that relates the dipole and the displacement fields
\begin{equation}
	\B {\C P} =-\kappa^2 \B d\ , \quad (\text{with only one screening parameter}) \ .
	\label{Pvsd}
\end{equation}
This constitutive relation means that the dipole and the displacement fields are expected to be co-linear and opposite in direction. This prediction was not corroborated in simulations, an angle was distinctly existing between this prediction and the measured direction of the dipole field \cite{23MMPR}. To come to grip with the data, one needs to allow a more general tensor $\B \Gamma$ \cite{tuhin2024,fu2024odddipole}, 
\begin{equation}
 \B \Gamma = \begin{pmatrix}
\kappa^2_e & \mp\kappa^2_o \\
\pm\kappa^2_o & \kappa^2_e
\end{pmatrix} \ .
\label{odd0}
\end{equation}
The consequence of using this matrix in Eq.~(\ref{d2}) is a change in the constitutive relation (\ref{Pvsd}). Now there is an angle between the dipole and the displacement vectors, and the angle is $\theta_\kappa$, satisfying 
$ \tan \theta_\kappa = \pm\frac{\kappa^2_o}{\kappa^2_e}$. For any given
sample the angle can be either positive or negative, depending on the sign before $\kappa_0^2$, exhibiting Chiral symmetry breaking. 

While the relevance of Eqs.~(\ref{d2}) and (\ref{odd0}) to various examples of straining protocols on amorphous solids was demonstrated recently \cite{22MMPRSZ,24JPS,fu2024odddipole,tuhin2024}, including careful comparison of its predictions to both experiments and simulations, what is the {\em selection principle} of the emergent screening parameters $\kappa_e$ and $\kappa_o$, (if they exist),  was not studied in full. This Letter aims to elaborate on such a principle. To this aim, we re-consider two straining protocols for which Eqs.~(\ref{d2}) and (\ref{odd0}) can be solved analytically. These two protocols are the flattening of a spherical cap and the inflation of a central hole. The
Numerical simulation details are presented in the section Methods at the end of the Letter.

\begin{figure}
	\includegraphics[width=0.65\linewidth]{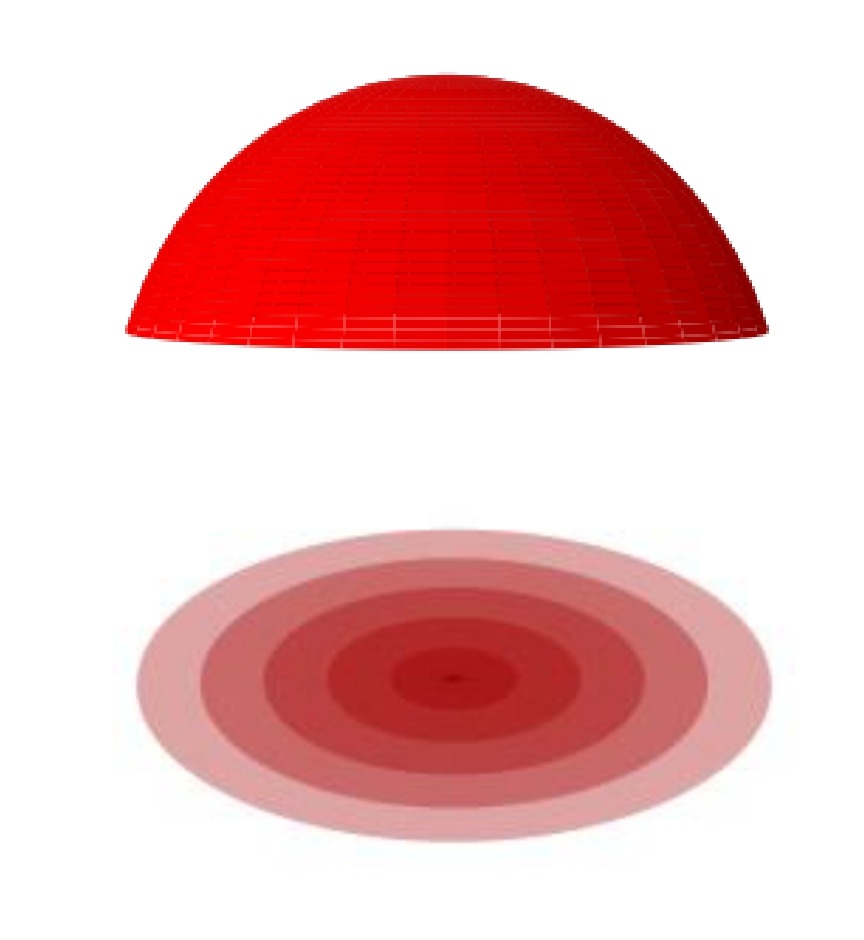}
	\caption{The spherical cap is flattened onto the plane} 
	\label{flattening}
\end{figure}

{\bf First protocol: flattening of a spherical cap:} in the first protocol we explore the response of a 2-dimensional sheet of amorphous solid in the form of a cap, that is flattened onto a plane, see Fig.~\ref{flattening}.  The advantage of this mode of loading is that it can also be studied analytically and by numerical simulations, offering a clear-cut example of the role of plastic deformations in introducing screening to the elastic response of amorphous solids.
\begin{figure}
	\includegraphics[width=0.8\linewidth]{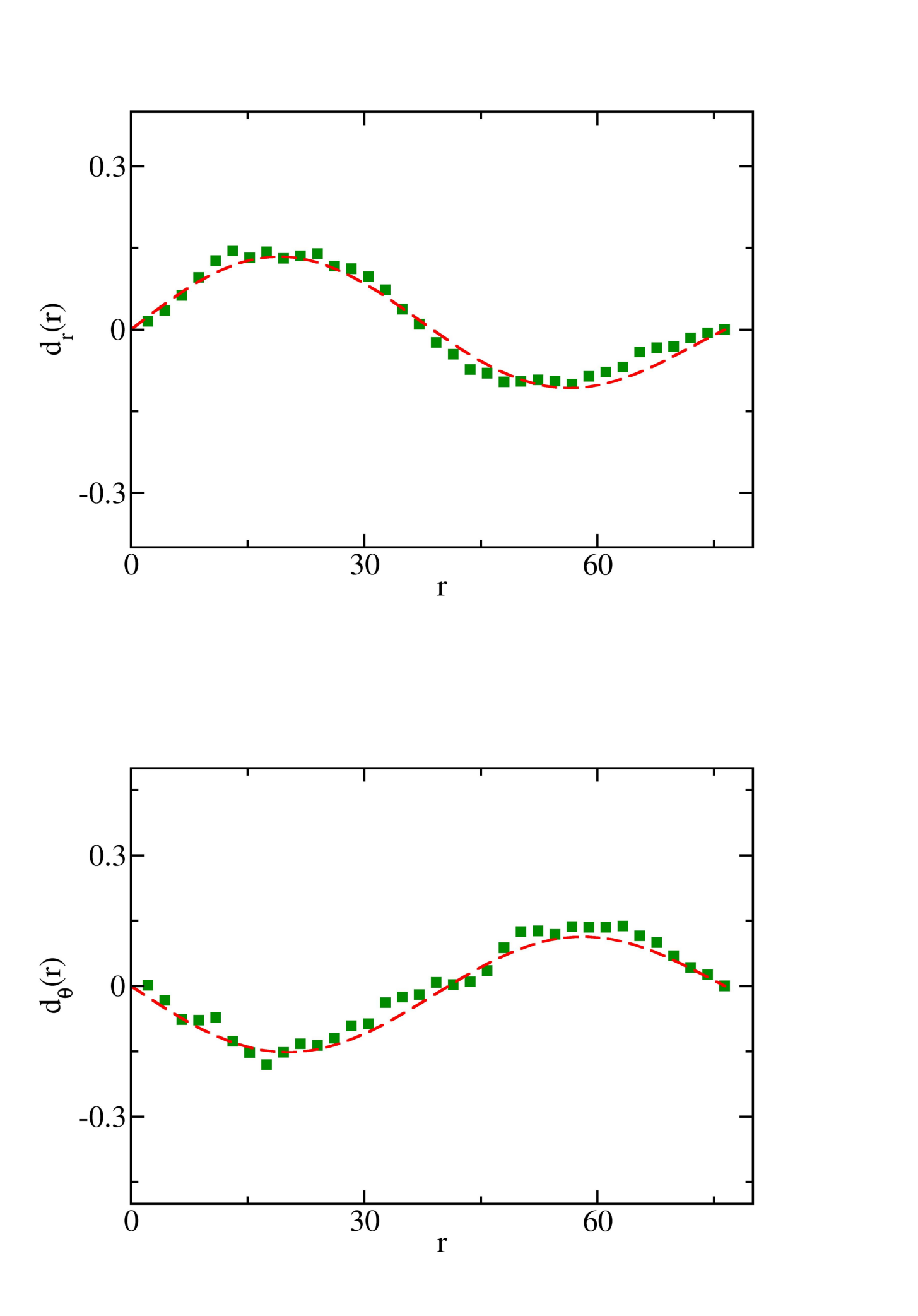}
	\caption{ A typical example of the angle-averaged radial and tangential components of the displacement field for flattening a spherical cap with a fixed imposed radius of curvature, $R = 500$. The green squares represent the results of numerical simulations, which are averages calculated over five realizations under identical control parameters. The red dashed line is the theoretical fit. The fitted values of screening parameters are $\kappa_e$ = 0.014 and $\kappa_o$ =0.0063. }
	\label{radialtrans}
\end{figure}
To allow both the analytic and the numerical investigation we need to clarify first the mechanical loading that the process shown in Fig.~\ref{flattening} implies. To this aim, we will consider a flat circular piece of material of radius $r_{\rm out}$ and ask how to inflate each area element in order to load the material precisely as if it was flattened from the spherical cap. Employing a local inflation factor $\exp[2\phi(\B x)]$, the resulting Gaussian curvature is \cite{tuhin2024}
\begin{equation}
	K_g \equiv -e^{-2\phi(\B x)}\Delta \phi(\B x) \ , 
\end{equation}
with $\Delta$ being the usual Laplace operator. To respect the constant Gaussian curvature of the spherical cap we demand $K_g$ to be constant, say $\alpha$. The solution of the equation $K_g=\alpha$ which is finite at the origin reads 
\begin{equation}
	e^{2\phi(\B x)}=\frac{4c^2}{(\alpha r^2 +c^2)^2} \ .
	\label{inflate}
\end{equation}
Here, $\alpha$ = 1/$R^2$ is the Gaussian curvature of our parameter space for the numerical simulations. In these simulations, after the differential inflation implied by Eq.~(\ref{inflate}), one performs energy gradient minimization until the forces on each particle vanish (up to a tolerance of $10^{-6}$). The resulting displacement field $\B d(r,\theta)$ is then measured (with $r$ and $\theta$ being the polar coordinates). Finally, both the radial and the transverse components of the displacement field  are obtained as an angle average, 
\begin{eqnarray}
	d_r(r)&\equiv& (2\pi)^{-1} \oint_{0}^{2\pi} \B d(r,\theta)\cdot \hat{r} d\theta \ , \nonumber\\
	d_\theta(r) &\equiv& (2\pi)^{-1} \oint_{0}^{2\pi} \B d(r,\theta)\cdot \hat{\theta} d\theta  \ .
	\label{components}
\end{eqnarray}
Here $\hat r\equiv \B r/r$ and $\hat \theta$ is its orthogonal unit vector.
The analytic solutions of these functions under this flattening protocol were presented in full detail in Ref~\cite{tuhin2024}. 

A typical comparison of the numerically found solutions and the analytic fits are shown in Fig.~\ref{radialtrans}. The parameters used for this include a 2D Poisson's ratio  (\(\nu = 0.153\)) and a radius of curvature (\(R = 500\)).
For the present example, the best fits required the values $\kappa_e$ = 0.014 and $\kappa_o$ =0.0063 for the screening parameters. With other choices of geometry and radius of curvature, other values of screening parameters are found. The question is why these values
emerge, and can we predict them a-priori. The positive answers are provided below.

\begin{figure}
	\includegraphics[width=0.25\textwidth]{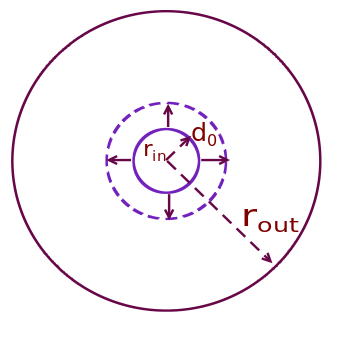}
	\caption{Schematic diagram of the inflation of a central hole. Material is confined in an annulus between an outer disk of radius $r_{\rm out}$ and an inner boundary of radius $r_{\rm in}$. The inner radius is then inflated to $r_{\rm in}+d_0$ and then the material is relaxed back to mechanical equilibrium.}
	\label{inflation}
\end{figure}

{\bf Second protocol: inflation of central hole:} This protocol was studied extensively in simulations in two and three dimensions, and also in an experiment \cite{21LMMPRS,22MMPRSZ,22BMP,22KMPS,23CMP}. We will focus here on two-dimensional simulations, employing point particles with Lennard-Jones interactions, prepared with a desired target pressure $P$ and confined in a circular two-dimensional area with a fixed outer circular wall of radius $r_{\rm out}$.
One disk of radius $r_{\rm in}$ is fixed to the center, and after equilibration it is inflated by a chosen amount $d_0$, $r_{\rm in}\to r_{\rm in}+d_0 $. This straining protocol is shown in Fig.~\ref{inflation}. 
After energy minimization, the resulting displacement field $\B d(r,\theta)$  is measured; as before, both the radial and the transverse components of the displacement field are obtained as the angle average defined by Eqs. (\ref{components}). 

\begin{figure}
	\includegraphics[width=0.40\textwidth]{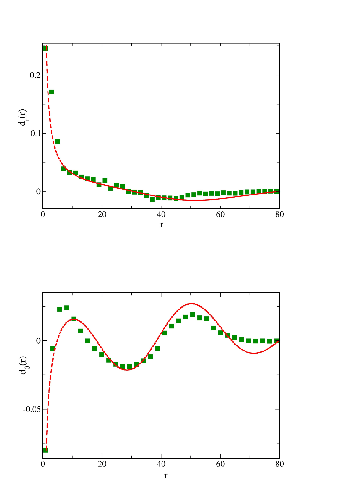}
	\caption{Comparison of measured angle averaged radial, $d_r(r)$ (upper panel) and transverse, $d_{\theta}(r)$ (lower panel) components of displacement field to the theory (dashed line) with $\kappa_e = 0.164$, $\kappa_o=0.128$ and $\tilde{\lambda} = 7.5$. The displacement fields result from inflating the inner radius $r_{in} = 1.2$ of the system by $d_0 =0.25 $, where $r_{out} = 80$ and $d_{\theta}(r_{in}) = -0.08$.}
	\label{GlassFit}
\end{figure}

A typical comparison of the numerically found solutions and the analytic fits is shown in Fig.~\ref{GlassFit}. In this case, the data represent an ensemble average over five realizations with the same control parameters. 
 For the present example, the best fits required the values $\kappa_e$ = 0.164 and $\kappa_o$ =0.128 for the screening parameters. As in the first example, , choices of other geometry and inflation $d_0$ choices lead to other values of screening parameters. The same question of why these values
emerge is discussed next.

\begin{figure}
	\includegraphics[width=0.50\textwidth]{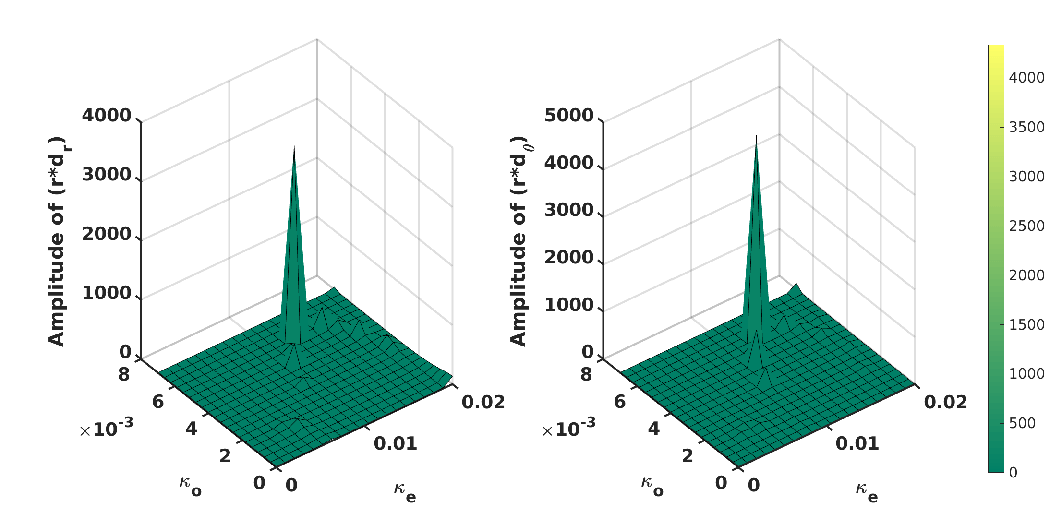}
\caption{The theoretical predictions for the maximal amplitude of $r\times d_r(r)$ and $r\times d_\theta(r)$ in the flattening protocol. The functions are plotted using Eq.~(C2) from the Ref~\cite{tuhin2024}.} 
\label{singflat}
\end{figure}

{\bf Selection Principle:} To understand how the emergent values of the screening parameters are selected, we need to examine closely the analytic solutions for the functions appearing in Eqs.~\ref{components}. These are presented in full detail in Refs.\cite{tuhin2024,fu2024odddipole} for the flattening and inflation protocols, respectively, and there is no reason to reproduce them here in any detail. It suffices to recognize that these functions depend explicitly on the screening parameters $\kappa_e$ and $\kappa_o$, and it makes sense to designate this explicitly, renaming the functions as
$d_r(r; \kappa_e,\kappa_o)$ and $d_\theta(r; \kappa_e,\kappa_o)$. In fact, these functions depend {\em sensitively} on the screening parameters, exhibiting singular behavior near certain values of these. As an example, we present in Fig.~\ref{singflat} the maximal amplitude of $r\times d_r(r; \kappa_e,\kappa_o)$ and $r \times d_\theta(r; \kappa_e,\kappa_o)$ for the geometric parameters employed to produce Fig.~\ref{radialtrans}. We multiply by $r$ to compensate for any $1/r$ power-law decay.
We discover that for the flattening protocols, there exist preferred values of $\kappa_e$ and $\kappa_o$ that maximize the anomalous response of the system. Moreover, these values are the same for the two functions, $d_r(r; \kappa_e,\kappa_o)$ and $d_\theta(r; \kappa_e,\kappa_o)$. These values are
 $\kappa_e=0.013, \kappa_o= 0.006)$. These values should be compared to the best fit emergent parameters as seen in Fig.~\ref{radialtrans}, i.e.  $\kappa_e=0.014, \kappa_o= 0.0063)$. Needless to say, we checked that this agreement is not accidental, and it is consistent in all the simulations that we performed. 
 
\begin{figure}
 	\includegraphics[width=0.40\textwidth]{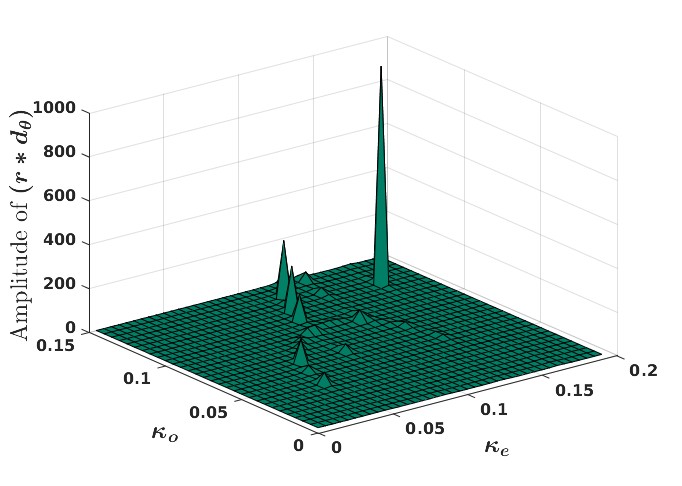}
 	\caption{The theoretical predictions for the maximal amplitude of $r\times d_\theta(r)$ in the inflation protocol. The functions are plotted using Eqs.~(A29,A30) from the Ref~\cite{fu2024odddipole}.} 
 	\label{singinflate}
 \end{figure}

 A similar selection principle operates in the inflation protocol. In this case the functions $d_r(r; \kappa_e,\kappa_o)$ and $d_\theta(r; \kappa_e,\kappa_o)$ are presented in detail in Ref.~\cite{fu2024odddipole} and for the geometric parameters pertaining to the data in Fig.~\ref{GlassFit}, the function $r\times d_\theta(r)$ is plotted in Fig.~\ref{singinflate}. There are a number of peaklets, but the most
 prominent peak identifies the screening parameters  $\kappa_e=0.17, \kappa_o= 0.125$. We should compare these apriori predicted values with the actual best fits presented in Fig.~\ref{GlassFit}, i.e. $\kappa_e=0.164, \kappa_o= 0.128$. Also, for the inflation protocols, we have ascertained that similarly close predictions are consistently found for other geometric parameters and inflation amplitudes. 
 
 {\bf Conclusion and summary :}
The aim of this Letter was to examine why certain values of screening parameters emerge as a result of screening by dipole charges in the displacement field that follows a nonuniform mechanical strain in amorphous solids. We find that these
screening parameters are not random, they are direct
consequences of the system’s geometry, boundary conditions and applied strain. We discover that the mechanical response is most intense at discrete values of
screening parameters, and when the geometry of the system allows these particular values of inverse length scales,
they emerge spontaneously as the preferred ones. Therefore, by examining the theoretical solutions for the displacement field, which in the present cases are analytic,
one can predict a-priori which screening parameters will
emerge in a given context. By examining two very different protocols of strain, and finding similar characteristics,
we propose that this is a generic observation. Nevertheless, additional protocols, both experimental and simulational, are called for to support or delineate this important conclusion. Finally, searching for similar conclusions
in three-dimensional system is on our agenda for the near
future.

 {\bf Methods}  
 
 {\em Flattening protocol}
 
 Open-source codes (LAMMPS \cite{95Pli}) are used to perform the simulations. 
 For the flattening protocol, we employed four sizes of frictionless disks, of radii 0.4,0.5,0.6, and 0.7, placed randomly in a circular disk of radius $r_{\rm out}$. The simulations are performed with a total of 15,000 discs. An initial area fraction below the jamming threshold was chosen, and then the outer radius was isotropically reduced to achieve a finite chosen pressure. As before, in each step, energy was minimized. This process is carried out until the desired target pressure is reached and forces are minimized to values smaller than $10^{-6}$. The straining protocol then inflates each disk according to Eq.~(\ref{inflate}), and subsequently, mechanical equilibration follows, and
 the resulting displacement field is measured. In these simulations, $K_n=2000$. 
 The mass of each disk is $1$.

 {\em Inflation protocol}: To obtain the glass configurations, we simulate $N = 20,000$ poly-disperse point particles in an annulus with outer radius $r_{out}$ and inner bounday at $r_{in}$, such that density of the system confined in $A = \pi(r_{out}^2-r_{in}^2)$ is: $\rho = N/A = 1.0$. The binary interactions between point particles of mass $m=1$ is given as:
 
 \begin{equation}
 	U_{ij}(r) = 
 	\begin{cases}
 		u_{ij}^{LJ} + A_{ij} + B_{ij}r + C_{ij}r^2 \;\; \textbf{if} \;r_{ij} \leq r_{cut} \\
 		0, \hskip 40mm \textbf{otherwise}
 	\end{cases}
 	\label{Eq1}
 \end{equation}	
 
 where $u_{ij}^{LJ}$ is standard Lennard-Jones potential given in Eq. \ref{LJ} and $A_{ij} = 0.4526\epsilon_{ij}$, $B_{ij} = -3100\epsilon_{ij}/\sigma_{ij}$, $C_{ij} = 0.0542\epsilon_{ij}/\sigma_{ij}^2$ are added to smooth the potential at cuf-off distance, $r_{cut}=2.5\sigma_{ij}$ (upto second derivative)\\
 \begin{equation}
 	u_{ij}^{LJ} = 4\epsilon \Big[ \Big(\frac{\sigma_{ij}}{r}\Big)^{12} - \Big(\frac{\sigma_{ij}}{r}\Big)^{6} \Big]
 	\label{LJ}
 \end{equation}
 
 The interaction length for each particle, $\sigma_i$ is drawn from a probability distribution: $P(\sigma) \simeq 1/\sigma^3$ in a range between $\sigma_{max} = 1.61$ and $\sigma_{min}=\sigma_{max}/2.219$ with mean, $\bar{\sigma} =1$. The mixing rule of $\sigma$ for interparticle interactions is:
 
 \begin{equation}
 	\sigma_{ij}= \frac{\sigma_i + \sigma_j}{2}\Big[ 1-0.2|\sigma_i -\sigma_j| \Big]
 \end{equation}

 The system is thermalized at mother temperature, $T_m=1$ using the swap Monte-Carlo method, and then cooled down to $T=0$ using the conjugate gradient method. Once the system achieves mechanical equilibrium such that force on each particle is less than $10^{-8}$, we inflate the inner radius by $d_0$, as shown in Fig.~\ref{inflation}.

\bibliography{ALL.anomalous}

\begin{thebibliography}{23}%
\makeatletter
\providecommand \@ifxundefined [1]{%
 \@ifx{#1\undefined}
}%
\providecommand \@ifnum [1]{%
 \ifnum #1\expandafter \@firstoftwo
 \else \expandafter \@secondoftwo
 \fi
}%
\providecommand \@ifx [1]{%
 \ifx #1\expandafter \@firstoftwo
 \else \expandafter \@secondoftwo
 \fi
}%
\providecommand \natexlab [1]{#1}%
\providecommand \enquote  [1]{``#1''}%
\providecommand \bibnamefont  [1]{#1}%
\providecommand \bibfnamefont [1]{#1}%
\providecommand \citenamefont [1]{#1}%
\providecommand \href@noop [0]{\@secondoftwo}%
\providecommand \href [0]{\begingroup \@sanitize@url \@href}%
\providecommand \@href[1]{\@@startlink{#1}\@@href}%
\providecommand \@@href[1]{\endgroup#1\@@endlink}%
\providecommand \@sanitize@url [0]{\catcode `\\12\catcode `\$12\catcode `\&12\catcode `\#12\catcode `\^12\catcode `\_12\catcode `\%12\relax}%
\providecommand \@@startlink[1]{}%
\providecommand \@@endlink[0]{}%
\providecommand \url  [0]{\begingroup\@sanitize@url \@url }%
\providecommand \@url [1]{\endgroup\@href {#1}{\urlprefix }}%
\providecommand \urlprefix  [0]{URL }%
\providecommand \Eprint [0]{\href }%
\providecommand \doibase [0]{https://doi.org/}%
\providecommand \selectlanguage [0]{\@gobble}%
\providecommand \bibinfo  [0]{\@secondoftwo}%
\providecommand \bibfield  [0]{\@secondoftwo}%
\providecommand \translation [1]{[#1]}%
\providecommand \BibitemOpen [0]{}%
\providecommand \bibitemStop [0]{}%
\providecommand \bibitemNoStop [0]{.\EOS\space}%
\providecommand \EOS [0]{\spacefactor3000\relax}%
\providecommand \BibitemShut  [1]{\csname bibitem#1\endcsname}%
\let\auto@bib@innerbib\@empty
\bibitem [{\citenamefont {Zallen}(2008)}]{08Zal}%
  \BibitemOpen
  \bibfield  {author} {\bibinfo {author} {\bibfnamefont {R.}~\bibnamefont {Zallen}},\ }\href@noop {} {\emph {\bibinfo {title} {The physics of amorphous solids}}}\ (\bibinfo  {publisher} {John Wiley \& Sons},\ \bibinfo {year} {2008})\BibitemShut {NoStop}%
\bibitem [{\citenamefont {Alexander}(1998)}]{98Ale}%
  \BibitemOpen
  \bibfield  {author} {\bibinfo {author} {\bibfnamefont {S.}~\bibnamefont {Alexander}},\ }\bibfield  {title} {\bibinfo {title} {Amorphous solids: their structure, lattice dynamics and elasticity},\ }\href@noop {} {\bibfield  {journal} {\bibinfo  {journal} {Physics reports}\ }\textbf {\bibinfo {volume} {296}},\ \bibinfo {pages} {65} (\bibinfo {year} {1998})}\BibitemShut {NoStop}%
\bibitem [{\citenamefont {Jaeger}\ \emph {et~al.}(1996)\citenamefont {Jaeger}, \citenamefont {Nagel},\ and\ \citenamefont {Behringer}}]{jaeger1996granular}%
  \BibitemOpen
  \bibfield  {author} {\bibinfo {author} {\bibfnamefont {H.~M.}\ \bibnamefont {Jaeger}}, \bibinfo {author} {\bibfnamefont {S.~R.}\ \bibnamefont {Nagel}},\ and\ \bibinfo {author} {\bibfnamefont {R.~P.}\ \bibnamefont {Behringer}},\ }\bibfield  {title} {\bibinfo {title} {Granular solids, liquids, and gases},\ }\href@noop {} {\bibfield  {journal} {\bibinfo  {journal} {Reviews of modern physics}\ }\textbf {\bibinfo {volume} {68}},\ \bibinfo {pages} {1259} (\bibinfo {year} {1996})}\BibitemShut {NoStop}%
\bibitem [{\citenamefont {Bera}\ \emph {et~al.}(2024)\citenamefont {Bera}, \citenamefont {Baggioli}, \citenamefont {Petersen}, \citenamefont {Sirk}, \citenamefont {Liu},\ and\ \citenamefont {Zaccone}}]{10.1093/pnasnexus/pgae315}%
  \BibitemOpen
  \bibfield  {author} {\bibinfo {author} {\bibfnamefont {A.}~\bibnamefont {Bera}}, \bibinfo {author} {\bibfnamefont {M.}~\bibnamefont {Baggioli}}, \bibinfo {author} {\bibfnamefont {T.~C.}\ \bibnamefont {Petersen}}, \bibinfo {author} {\bibfnamefont {T.~W.}\ \bibnamefont {Sirk}}, \bibinfo {author} {\bibfnamefont {A.~C.~Y.}\ \bibnamefont {Liu}},\ and\ \bibinfo {author} {\bibfnamefont {A.}~\bibnamefont {Zaccone}},\ }\bibfield  {title} {\bibinfo {title} {{Clustering of negative topological charges precedes plastic failure in 3D glasses}},\ }\href {https://doi.org/10.1093/pnasnexus/pgae315} {\bibfield  {journal} {\bibinfo  {journal} {PNAS Nexus}\ }\textbf {\bibinfo {volume} {3}},\ \bibinfo {pages} {pgae315} (\bibinfo {year} {2024})}\BibitemShut {NoStop}%
\bibitem [{\citenamefont {Blumenfeld}(2004)}]{BlumPRL}%
  \BibitemOpen
  \bibfield  {author} {\bibinfo {author} {\bibfnamefont {R.}~\bibnamefont {Blumenfeld}},\ }\bibfield  {title} {\bibinfo {title} {Stresses in isostatic granular systems and emergence of force chains},\ }\href {https://doi.org/10.1103/PhysRevLett.93.108301} {\bibfield  {journal} {\bibinfo  {journal} {Phys. Rev. Lett.}\ }\textbf {\bibinfo {volume} {93}},\ \bibinfo {pages} {108301} (\bibinfo {year} {2004})}\BibitemShut {NoStop}%
\bibitem [{\citenamefont {Liu}\ and\ \citenamefont {Nagel}(1993)}]{NagelPRB}%
  \BibitemOpen
  \bibfield  {author} {\bibinfo {author} {\bibfnamefont {C.-h.}\ \bibnamefont {Liu}}\ and\ \bibinfo {author} {\bibfnamefont {S.~R.}\ \bibnamefont {Nagel}},\ }\bibfield  {title} {\bibinfo {title} {Sound in a granular material: Disorder and nonlinearity},\ }\href {https://doi.org/10.1103/PhysRevB.48.15646} {\bibfield  {journal} {\bibinfo  {journal} {Phys. Rev. B}\ }\textbf {\bibinfo {volume} {48}},\ \bibinfo {pages} {15646} (\bibinfo {year} {1993})}\BibitemShut {NoStop}%
\bibitem [{\citenamefont {Landau}\ and\ \citenamefont {Lifshitz}(1970)}]{Landau}%
  \BibitemOpen
  \bibfield  {author} {\bibinfo {author} {\bibfnamefont {L.~D.}\ \bibnamefont {Landau}}\ and\ \bibinfo {author} {\bibfnamefont {E.~M.}\ \bibnamefont {Lifshitz}},\ }\bibfield  {title} {\bibinfo {title} {{Theory of Elasticity}},\ }\bibfield  {journal} {\bibinfo  {journal} {Course of Theoretical Physics}\ }\href {https://doi.org/10.1126/science.1070375} {10.1126/science.1070375} (\bibinfo {year} {1970})\BibitemShut {NoStop}%
\bibitem [{\citenamefont {Karmakar}\ \emph {et~al.}(2010)\citenamefont {Karmakar}, \citenamefont {Lerner},\ and\ \citenamefont {Procaccia}}]{10KLP}%
  \BibitemOpen
  \bibfield  {author} {\bibinfo {author} {\bibfnamefont {S.}~\bibnamefont {Karmakar}}, \bibinfo {author} {\bibfnamefont {E.}~\bibnamefont {Lerner}},\ and\ \bibinfo {author} {\bibfnamefont {I.}~\bibnamefont {Procaccia}},\ }\bibfield  {title} {\bibinfo {title} {{Athermal nonlinear elastic constants of amorphous solids}},\ }\href {https://doi.org/10.1103/PhysRevE.82.026105} {\bibfield  {journal} {\bibinfo  {journal} {Phys. Rev.E}\ }\textbf {\bibinfo {volume} {82}},\ \bibinfo {pages} {026105} (\bibinfo {year} {2010})}\BibitemShut {NoStop}%
\bibitem [{\citenamefont {Hentschel}\ \emph {et~al.}(2011)\citenamefont {Hentschel}, \citenamefont {Karmakar}, \citenamefont {Lerner},\ and\ \citenamefont {Procaccia}}]{11HKLP}%
  \BibitemOpen
  \bibfield  {author} {\bibinfo {author} {\bibfnamefont {H.~G.~E.}\ \bibnamefont {Hentschel}}, \bibinfo {author} {\bibfnamefont {S.}~\bibnamefont {Karmakar}}, \bibinfo {author} {\bibfnamefont {E.}~\bibnamefont {Lerner}},\ and\ \bibinfo {author} {\bibfnamefont {I.}~\bibnamefont {Procaccia}},\ }\bibfield  {title} {\bibinfo {title} {{Do athermal amorphous solids exist?}},\ }\href {https://doi.org/10.1103/PhysRevE.83.061101} {\bibfield  {journal} {\bibinfo  {journal} {Phys. Rev.E}\ }\textbf {\bibinfo {volume} {83}},\ \bibinfo {pages} {061101} (\bibinfo {year} {2011})}\BibitemShut {NoStop}%
\bibitem [{\citenamefont {Eshelby}(1957)}]{54Esh}%
  \BibitemOpen
  \bibfield  {author} {\bibinfo {author} {\bibfnamefont {J.~D.}\ \bibnamefont {Eshelby}},\ }\bibfield  {title} {\bibinfo {title} {The determination of the elastic field of an ellipsoidal inclusion, and related problems},\ }\href {https://doi.org/10.1098/rspa.1957.0133} {\bibfield  {journal} {\bibinfo  {journal} {Proceedings of the Royal Society of London A: Mathematical, Physical and Engineering Sciences}\ }\textbf {\bibinfo {volume} {241}},\ \bibinfo {pages} {376} (\bibinfo {year} {1957})}\BibitemShut {NoStop}%
\bibitem [{\citenamefont {Lema\^{\i}tre}\ \emph {et~al.}(2021)\citenamefont {Lema\^{\i}tre}, \citenamefont {Mondal}, \citenamefont {Moshe}, \citenamefont {Procaccia}, \citenamefont {Roy},\ and\ \citenamefont {Screiber-Re'em}}]{21LMMPRS}%
  \BibitemOpen
  \bibfield  {author} {\bibinfo {author} {\bibfnamefont {A.}~\bibnamefont {Lema\^{\i}tre}}, \bibinfo {author} {\bibfnamefont {C.}~\bibnamefont {Mondal}}, \bibinfo {author} {\bibfnamefont {M.}~\bibnamefont {Moshe}}, \bibinfo {author} {\bibfnamefont {I.}~\bibnamefont {Procaccia}}, \bibinfo {author} {\bibfnamefont {S.}~\bibnamefont {Roy}},\ and\ \bibinfo {author} {\bibfnamefont {K.}~\bibnamefont {Screiber-Re'em}},\ }\bibfield  {title} {\bibinfo {title} {Anomalous elasticity and plastic screening in amorphous solids},\ }\href {https://doi.org/10.1103/PhysRevE.104.024904} {\bibfield  {journal} {\bibinfo  {journal} {Phys. Rev. E}\ }\textbf {\bibinfo {volume} {104}},\ \bibinfo {pages} {024904} (\bibinfo {year} {2021})}\BibitemShut {NoStop}%
\bibitem [{\citenamefont {Mondal}\ \emph {et~al.}(2022)\citenamefont {Mondal}, \citenamefont {Moshe}, \citenamefont {Procaccia}, \citenamefont {Roy}, \citenamefont {Shang},\ and\ \citenamefont {Zhang}}]{22MMPRSZ}%
  \BibitemOpen
  \bibfield  {author} {\bibinfo {author} {\bibfnamefont {C.}~\bibnamefont {Mondal}}, \bibinfo {author} {\bibfnamefont {M.}~\bibnamefont {Moshe}}, \bibinfo {author} {\bibfnamefont {I.}~\bibnamefont {Procaccia}}, \bibinfo {author} {\bibfnamefont {S.}~\bibnamefont {Roy}}, \bibinfo {author} {\bibfnamefont {J.}~\bibnamefont {Shang}},\ and\ \bibinfo {author} {\bibfnamefont {J.}~\bibnamefont {Zhang}},\ }\bibfield  {title} {\bibinfo {title} {Experimental and numerical verification of anomalous screening theory in granular matter},\ }\href {https://doi.org/https://doi.org/10.1016/j.chaos.2022.112609} {\bibfield  {journal} {\bibinfo  {journal} {Chaos, Solitons and Fractals}\ }\textbf {\bibinfo {volume} {164}},\ \bibinfo {pages} {112609} (\bibinfo {year} {2022})}\BibitemShut {NoStop}%
\bibitem [{\citenamefont {Bhowmik}\ \emph {et~al.}(2022)\citenamefont {Bhowmik}, \citenamefont {Moshe},\ and\ \citenamefont {Procaccia}}]{22BMP}%
  \BibitemOpen
  \bibfield  {author} {\bibinfo {author} {\bibfnamefont {B.~P.}\ \bibnamefont {Bhowmik}}, \bibinfo {author} {\bibfnamefont {M.}~\bibnamefont {Moshe}},\ and\ \bibinfo {author} {\bibfnamefont {I.}~\bibnamefont {Procaccia}},\ }\bibfield  {title} {\bibinfo {title} {Direct measurement of dipoles in anomalous elasticity of amorphous solids},\ }\href {https://doi.org/10.1103/PhysRevE.105.L043001} {\bibfield  {journal} {\bibinfo  {journal} {Phys. Rev. E}\ }\textbf {\bibinfo {volume} {105}},\ \bibinfo {pages} {L043001} (\bibinfo {year} {2022})}\BibitemShut {NoStop}%
\bibitem [{\citenamefont {Kumar}\ \emph {et~al.}(2022)\citenamefont {Kumar}, \citenamefont {Moshe}, \citenamefont {Procaccia},\ and\ \citenamefont {Singh}}]{22KMPS}%
  \BibitemOpen
  \bibfield  {author} {\bibinfo {author} {\bibfnamefont {A.}~\bibnamefont {Kumar}}, \bibinfo {author} {\bibfnamefont {M.}~\bibnamefont {Moshe}}, \bibinfo {author} {\bibfnamefont {I.}~\bibnamefont {Procaccia}},\ and\ \bibinfo {author} {\bibfnamefont {M.}~\bibnamefont {Singh}},\ }\bibfield  {title} {\bibinfo {title} {Anomalous elasticity in classical glass formers},\ }\href {https://doi.org/10.1103/PhysRevE.106.015001} {\bibfield  {journal} {\bibinfo  {journal} {Phys. Rev. E}\ }\textbf {\bibinfo {volume} {106}},\ \bibinfo {pages} {015001} (\bibinfo {year} {2022})}\BibitemShut {NoStop}%
\bibitem [{\citenamefont {Charan}\ \emph {et~al.}(2023)\citenamefont {Charan}, \citenamefont {Moshe},\ and\ \citenamefont {Procaccia}}]{23CMP}%
  \BibitemOpen
  \bibfield  {author} {\bibinfo {author} {\bibfnamefont {H.}~\bibnamefont {Charan}}, \bibinfo {author} {\bibfnamefont {M.}~\bibnamefont {Moshe}},\ and\ \bibinfo {author} {\bibfnamefont {I.}~\bibnamefont {Procaccia}},\ }\bibfield  {title} {\bibinfo {title} {Anomalous elasticity and emergent dipole screening in three-dimensional amorphous solids},\ }\href {https://doi.org/10.1103/PhysRevE.107.055005} {\bibfield  {journal} {\bibinfo  {journal} {Phys. Rev. E}\ }\textbf {\bibinfo {volume} {107}},\ \bibinfo {pages} {055005} (\bibinfo {year} {2023})}\BibitemShut {NoStop}%
\bibitem [{\citenamefont {Jin}\ \emph {et~al.}(2024)\citenamefont {Jin}, \citenamefont {Procaccia},\ and\ \citenamefont {Samanta}}]{24JPS}%
  \BibitemOpen
  \bibfield  {author} {\bibinfo {author} {\bibfnamefont {Y.}~\bibnamefont {Jin}}, \bibinfo {author} {\bibfnamefont {I.}~\bibnamefont {Procaccia}},\ and\ \bibinfo {author} {\bibfnamefont {T.}~\bibnamefont {Samanta}},\ }\bibfield  {title} {\bibinfo {title} {Intermediate phase between jammed and unjammed amorphous solids},\ }\href {https://doi.org/10.1103/PhysRevE.109.014902} {\bibfield  {journal} {\bibinfo  {journal} {Phys. Rev. E}\ }\textbf {\bibinfo {volume} {109}},\ \bibinfo {pages} {014902} (\bibinfo {year} {2024})}\BibitemShut {NoStop}%
\bibitem [{\citenamefont {Nelson}\ and\ \citenamefont {Halperin}(1979)}]{79NH}%
  \BibitemOpen
  \bibfield  {author} {\bibinfo {author} {\bibfnamefont {D.~R.}\ \bibnamefont {Nelson}}\ and\ \bibinfo {author} {\bibfnamefont {B.~I.}\ \bibnamefont {Halperin}},\ }\bibfield  {title} {\bibinfo {title} {Dislocation-mediated melting in two dimensions},\ }\href {https://doi.org/10.1103/PhysRevB.19.2457} {\bibfield  {journal} {\bibinfo  {journal} {Phys. Rev. B}\ }\textbf {\bibinfo {volume} {19}},\ \bibinfo {pages} {2457} (\bibinfo {year} {1979})}\BibitemShut {NoStop}%
\bibitem [{\citenamefont {Zippelius}\ \emph {et~al.}(1980)\citenamefont {Zippelius}, \citenamefont {Halperin},\ and\ \citenamefont {Nelson}}]{80ZHN}%
  \BibitemOpen
  \bibfield  {author} {\bibinfo {author} {\bibfnamefont {A.}~\bibnamefont {Zippelius}}, \bibinfo {author} {\bibfnamefont {B.~I.}\ \bibnamefont {Halperin}},\ and\ \bibinfo {author} {\bibfnamefont {D.~R.}\ \bibnamefont {Nelson}},\ }\bibfield  {title} {\bibinfo {title} {Dynamics of two-dimensional melting},\ }\href {https://doi.org/10.1103/PhysRevB.22.2514} {\bibfield  {journal} {\bibinfo  {journal} {Phys. Rev. B}\ }\textbf {\bibinfo {volume} {22}},\ \bibinfo {pages} {2514} (\bibinfo {year} {1980})}\BibitemShut {NoStop}%
\bibitem [{\citenamefont {Kosterlitz}(2016)}]{16Kos}%
  \BibitemOpen
  \bibfield  {author} {\bibinfo {author} {\bibfnamefont {J.~M.}\ \bibnamefont {Kosterlitz}},\ }\bibfield  {title} {\bibinfo {title} {Kosterlitz{\textendash}thouless physics: a review of key issues},\ }\href {https://doi.org/10.1088/0034-4885/79/2/026001} {\bibfield  {journal} {\bibinfo  {journal} {Reports on Progress in Physics}\ }\textbf {\bibinfo {volume} {79}},\ \bibinfo {pages} {026001} (\bibinfo {year} {2016})}\BibitemShut {NoStop}%
\bibitem [{\citenamefont {Mondal}\ \emph {et~al.}(2023)\citenamefont {Mondal}, \citenamefont {Moshe}, \citenamefont {Procaccia},\ and\ \citenamefont {Roy}}]{23MMPR}%
  \BibitemOpen
  \bibfield  {author} {\bibinfo {author} {\bibfnamefont {C.}~\bibnamefont {Mondal}}, \bibinfo {author} {\bibfnamefont {M.}~\bibnamefont {Moshe}}, \bibinfo {author} {\bibfnamefont {I.}~\bibnamefont {Procaccia}},\ and\ \bibinfo {author} {\bibfnamefont {S.}~\bibnamefont {Roy}},\ }\href@noop {} {\bibinfo {title} {Dipole screening in pure shear strain protocols of amorphous solids}} (\bibinfo {year} {2023}),\ \Eprint {https://arxiv.org/abs/2305.11253} {arXiv:2305.11253 [cond-mat.soft Phys.Rev. E, in press]} \BibitemShut {NoStop}%
\bibitem [{\citenamefont {Livne}\ \emph {et~al.}(2024)\citenamefont {Livne}, \citenamefont {Samanta}, \citenamefont {Schiller}, \citenamefont {Procaccia},\ and\ \citenamefont {Moshe}}]{tuhin2024}%
  \BibitemOpen
  \bibfield  {author} {\bibinfo {author} {\bibfnamefont {N.~S.}\ \bibnamefont {Livne}}, \bibinfo {author} {\bibfnamefont {T.}~\bibnamefont {Samanta}}, \bibinfo {author} {\bibfnamefont {A.}~\bibnamefont {Schiller}}, \bibinfo {author} {\bibfnamefont {I.}~\bibnamefont {Procaccia}},\ and\ \bibinfo {author} {\bibfnamefont {M.}~\bibnamefont {Moshe}},\ }\href {https://arxiv.org/abs/2408.13086} {\bibinfo {title} {Continuum mechanics of differential growth in disordered granular matter}} (\bibinfo {year} {2024}),\ \Eprint {https://arxiv.org/abs/2408.13086} {arXiv:2408.13086 [cond-mat.soft]} \BibitemShut {NoStop}%
\bibitem [{\citenamefont {Fu}\ \emph {et~al.}(2024)\citenamefont {Fu}, \citenamefont {Hentschel}, \citenamefont {Kaur}, \citenamefont {Kumar},\ and\ \citenamefont {Procaccia}}]{fu2024odddipole}%
  \BibitemOpen
  \bibfield  {author} {\bibinfo {author} {\bibfnamefont {Y.}~\bibnamefont {Fu}}, \bibinfo {author} {\bibfnamefont {H.~G.~E.}\ \bibnamefont {Hentschel}}, \bibinfo {author} {\bibfnamefont {P.}~\bibnamefont {Kaur}}, \bibinfo {author} {\bibfnamefont {A.}~\bibnamefont {Kumar}},\ and\ \bibinfo {author} {\bibfnamefont {I.}~\bibnamefont {Procaccia}},\ }\href {https://arxiv.org/abs/2406.16075} {\bibinfo {title} {Odd dipole screening in radial inflation}} (\bibinfo {year} {2024}),\ \Eprint {https://arxiv.org/abs/2406.16075} {arXiv:2406.16075 [cond-mat.dis-nn]} \BibitemShut {NoStop}%
\bibitem [{\citenamefont {Plimpton}(1995)}]{95Pli}%
  \BibitemOpen
  \bibfield  {author} {\bibinfo {author} {\bibfnamefont {S.}~\bibnamefont {Plimpton}},\ }\bibfield  {title} {\bibinfo {title} {{Fast Parallel Algorithms for Short-Range Molecular Dynamics}},\ }\href {https://doi.org/10.1006/jcph.1995.1039} {\bibfield  {journal} {\bibinfo  {journal} {Journal of Computational Physics}\ }\textbf {\bibinfo {volume} {117}},\ \bibinfo {pages} {1} (\bibinfo {year} {1995})}\BibitemShut {NoStop}%
\end{thebibliography}%
\end{document}